# On the efficiency of the new Italian Senate and the role of 5 Stars Movement

## Comparison among different possible scenarios by means of a virtual Parliament model

A. Pluchino, A. Rapisarda, C. Garofalo, S. Spagano, M. Caserta
University of Catania, March 18th 2013

The recent **2013 Italian elections** are over and the situation that President Napolitano will have to settle soon for the formation of the new government is not the simplest one. After twenty years of bipolarism (more or less effective), where we were accustomed to a tight battle between two great political coalitions, the center-right and center-left, now, in the new Parliament, we have four political formations.

In addition to the two traditional coalitions led by **Democratic Party** and **Popolo della Libertà** (led by Berlusconi), we have seen the appearance of two other important parties: the one led by **Mario Monti,** called **Lista Civica,** and the other by **Beppe Grillo**, the **5 Stars Movement**. It was the strong affirmation of the latter, in particular, that changed the previous situation, bringing Italy in an apparent stalemate. In fact, thanks to the current (bad) electoral law, if in the Deputy Chamber the majority premium guarantees to the center-left coalition a large majority of deputies, in the Senate, no coalition alone possesses the absolute majority of seats and, to date, no one can see clear spaces for possible alliances leading to a stable government.

But is it really this result, as it would seem to suggest our common sense, the prelude to an inevitable phase of **ungovernability**? Can a Parliament with **changing majorities** in Senate to be as efficient as a Parliament with a large majority in both the Houses?

In this short note we will try to answer these questions going beyond common sense and analyzing the current political situation by means of a scientific, original and innovative instrument, i.e. an "**agent-based simulation**." Using a mathematical model that we have recently developed [1], we will try to quantify in numerical terms (within the context of a few, simple working hypothesis) the efficiency of the new Senate, comparing two possible scenarios which we might have. In particular, and this is our main result, we show that the situation is not so dramatic as it sounds, but it contains within itself potential positive aspects, as long as one makes the most appropriate choices.

In our model, **a Parliament can be simulated**, in a schematic way, as an aggregation of a number of deputies/senators, whose task is both to advance proposals of law and vote in favor or against those proposals (advanced by themselves or by other colleagues). In carrying out these actions, they may be motivated by self-interests, such as the re-election or other benefits, and / or by the general, collective benefit. Taking into account both of these factors, it is possible to represent the individual deputies/senators as points in a **diagram** made famous by the economist **C. Cipolla**, who, in 1976, introduced it in an essay on the fundamental laws of human stupidity. In our adaptation of the diagram to the political context, the x and y axes, whose range is in between [-1,1], represent, respectively, the degree of self-interest (personal gain) and collective interest (social gain) that lies behind the actions of each member of Parliament (MPs, see Fig.1). Obviously, people, and therefore MPs, do not always act in a consistent manner, but may behave differently depending on the circumstances. Thus, each point in the diagram represents a position representing an average over all the actions of a certain individual (we refer the reader to the bibliography for further details).

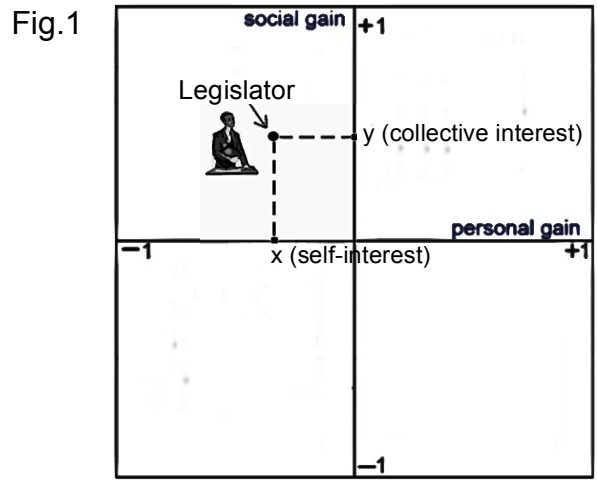

Fig.1

The **efficiency** of such a Parliament may, at this point, be calculated as the **product of the number of approved proposals times the collective advantage** resulting from them (i.e. the average value of the y coordinates of those laws). In our simulations, as already mentioned, we will consider only the Senate, as it came out of the Italian national elections in February 2013.

In the current situation we have **319 senators** distributed as follows: the center-left Coalition (**CSX**) won **123** seats, the center-right (**CDX**) **117** seats, 5 Stars Movement (**M5S**) **54** seats, the party led by **Monti 20** seats (19 + that one of Monti who is senator for life) and the remaining **5** seats (1 Vallee d'Aosta, 1st Movement Italians Abroad and three senators nominated by the President, called here "**independent**"). The simulations refer here to the two likely scenarios, assuming that some kind of government could get the **initial trust** (which is a fundamental step to start governing), and take into account the performance of a full legislature, during which a total of **1000** proposals will be advanced by the various senators.

These **two scenarios** are the following:

**- Scenario without alliances:** in this scenario there are no default alliances between parties and/or coalitions, then, after the initial trust, the Parliament work law after law for the whole term with variable majorities. In Figure 2 you can see two possible arrangements of CSX (in red), CDX (in blue), Monti's party (green) and 5 independent Senators (orange dots) on the Cipolla's diagram. As you can see, the senators members of parties or coalitions are within their "tolerance circles" (see bibliography). The **senators of M5S**, however, are represented by black dots distributed randomly and uniformly on the diagram, because their origin is extremely diverse and heterogeneous. In this regard, we have also considered **two possibilities**: the first takes into account the fact that the M5S is characterized by particular attention to the collective benefit of its proposals, therefore its senators are distributed only on the upper half-plane diagram (**M5S in the upper half-plane, Fig.2a**); the second, on the other hand, is more neutral and considers the senators distributed on the whole diagram (**M5S in the whole diagram, Fig.2b**).

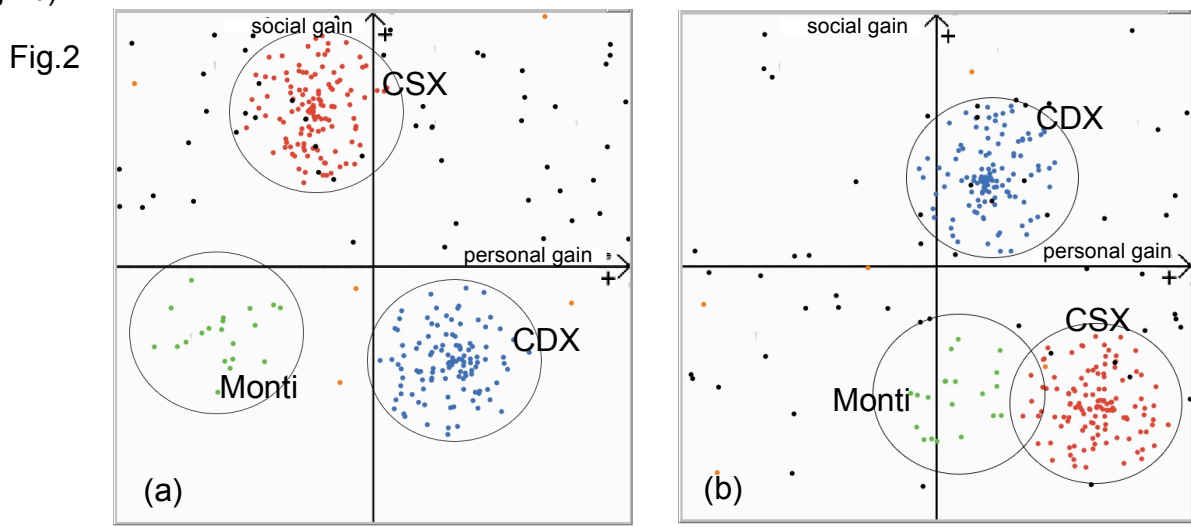

Fig.2

**- Scenario with "governissimo":** in this alternative scenario, the two center-right and center-left coalitions join their forces together with the Monti's party, forming a **governissimo** (i.e. a "big coalition" CSX + CDX + Monti) which, with its 260 seats, is well above the absolute majority of the Senate (160 seats), leaving the 5 Stars Movement and independents out of the government. In Fig.3 one can see two possible arrangements of this governissimo (whose senators lie into a single circle) and of the M5S senators that, as in the previous scenario, can occupy only the upper half-plane (Fig. 3a) or the entire Cipolla's diagram (Fig.3b).

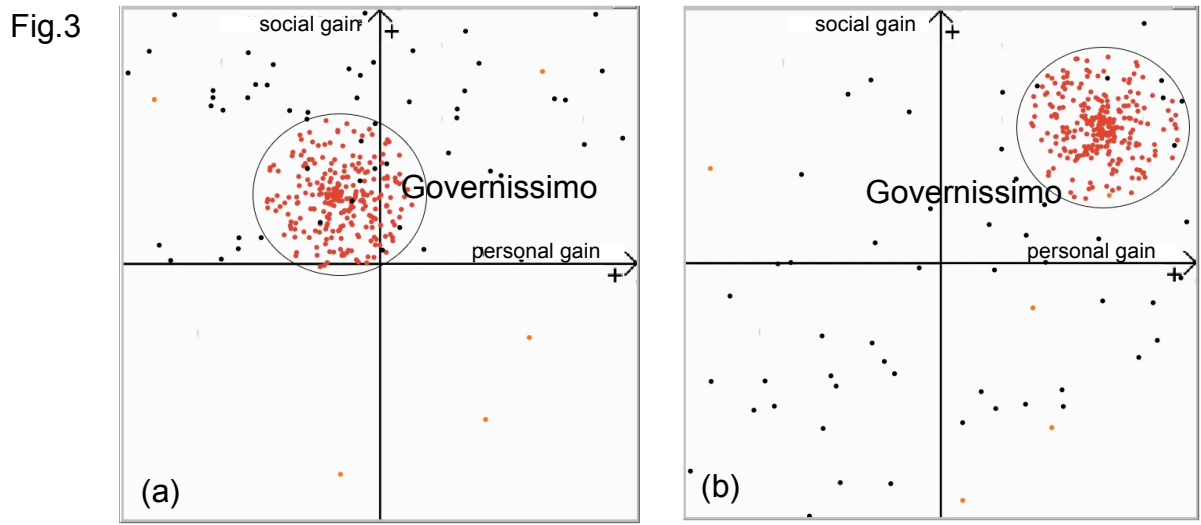

Fig.3

In our model, the **voting mechanism** is quite simple and schematic (again, for further details on the definition of "acceptance window" and "voting point", see the bibliography). The senators who are members of a political party or a coalition vote all together due to party discipline. In particular, they always vote in favor of the proposals put forward by any member of their own political party. On the other hand, for what concerns the proposals made by others, they vote according to the **"political center of gravity"** of their own party, i.e. the center of their own circle of tolerance in the Cipolla's diagram. The independents members however, in this case the five senators which do not belong to any political party, vote always in a free and uncoordinated way, regardless of who propose the bill. At this point, it is crucial the way in which the members of the **5 stars Movement** would vote. Actually, although this movement looks like a traditional political party, it is different not only for the selection of its members coming from very different social classes – this fact was already taken into account in the distribution of points over the Cipolla's diagram – but also for their internal organization, which is reflected in the voting strategy.

So we imagined **three possible ways** in which the senators of **M5S** could vote:

1) **Party discipline**: In this mode all the M5S senators **vote together**, as a traditional party. Their vote will always be in favor of their bills, while all the other proposals will be evaluated case by case, by comparing them with the position of the **fixed center of gravity** of their members in the Cipolla's diagram (which will be the point with coordinates (0, 0.5) in the hypothesis "M5S in the upper half-plane" case, or the point of coordinates (0.0) in the "M5S in the whole diagram" case);

2) **Web Discipline:** a characteristic feature of M5S is the interaction with their electorate through the web, so we figured that its senators can vote all together but **changing the center of gravity of the movement law by law**, according to the guidelines provided by the voters through web platforms such as, for example, "liquid feedback";

3) **Independent:** the senators M5S behave like **independent senators** and thus vote law by law independently of each other, without a binding mandate.

Taking into account all these variables and the different scenarios, we performed differrent sets of numerical simulations to quantify statistically the efficiency of the Senate. Each set included **100 independent legislatures**, each with a different random placement of both the senators and parties position on the Cipolla's diagram. From these sets, we extracted the **average efficiency** with an error of about 10% (taking into account also the possible changes in the range of the circle of tolerance of the parties). As a comparison, we remind that, in our previous study based on the introduction of randomly selected MPs in a bipolar Parliament [1], the peak of the maximum efficiency was always around a value of 15 (in arbitrary units).

Here below the tables and the graphical results of our simulations.

1) **Results with M5S in the upper half plane (Fig.4):** in this case, the "virtuous" location of M5S senators in the Cipolla's diagram yields beneficial effects, especially in the **scenario with variable majorities** (without alliances) and, in particular, when the voting of M5S follow the directives of the electorate in block **via the web**, a case which achieves an efficiency nearly twice (**12.7**) than the other ones. The scenario that sees the formation of a **governissimo** CSX CDX + Monti, instead, penalizes the efficiency of the Senate, which generally tends to decrease, remaining however positive thanks to the placement of M5S in the upper half plane. Finally, the "**independent**" voting mode of the M5S senators, while maintaining relatively a high efficiency (around 7), on the other hand tends to cancel the differences between the two scenarios.

| | M5S in the upper half plane | | |
|---|---|---|---|
| | PARTY DISCIPLINE | WEB DISCIPLINE | INDEPENDENT |
| Scenario without alliances | 6,1 | 12,7 | 7,5 |
| Scenario with "governissimo" | 3,1 | 5,5 | 6,8 |

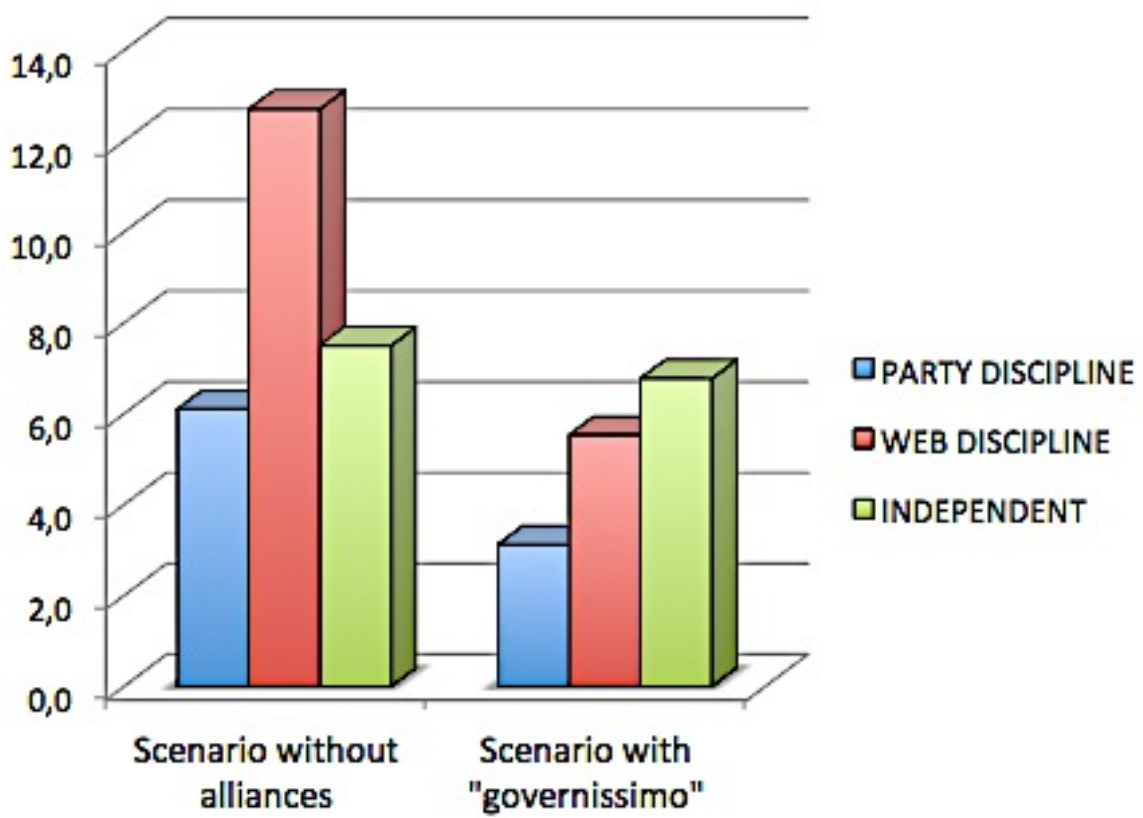

Fig.4

2) **Results with M5S in the whole diagram (Fig.5):** in this case we do not have the "virtuous" location of M5S in the Cipolla's diagram any longer and this fact has an important impact on the overall efficiency in all the simulations. In particular, we note a **significant difference between the results for the two scenarios** (without alliances and governissimo). In fact, while the scenario with variable majorities (without alliances) continues to have high efficiency values, especially in the two methods of voting for M5S senators which exclude the party discipline with a fixed center of gravity (around 9.5), on the other hand, the scenario with a governissimo comes out always severely penalized in terms of efficiency. We see that, in the latter case, the average efficiency becomes even negative with the two voting rules of M5S with fixed center of gravity, while it remains close to zero in the most favorable case of independent voting mode of the M5S senators.

| | M5S in the whole diagram | | |
| --- | --- | --- | --- |
| | PARTY DISCIPLINE | WEB DISCIPLINE | INDEPENDENT |
| Scenario without alliances | 4,3 | 9,2 | 9,8 |
| Scenario with "governissimo" | -4,2 | -2,8 | 0,3 |

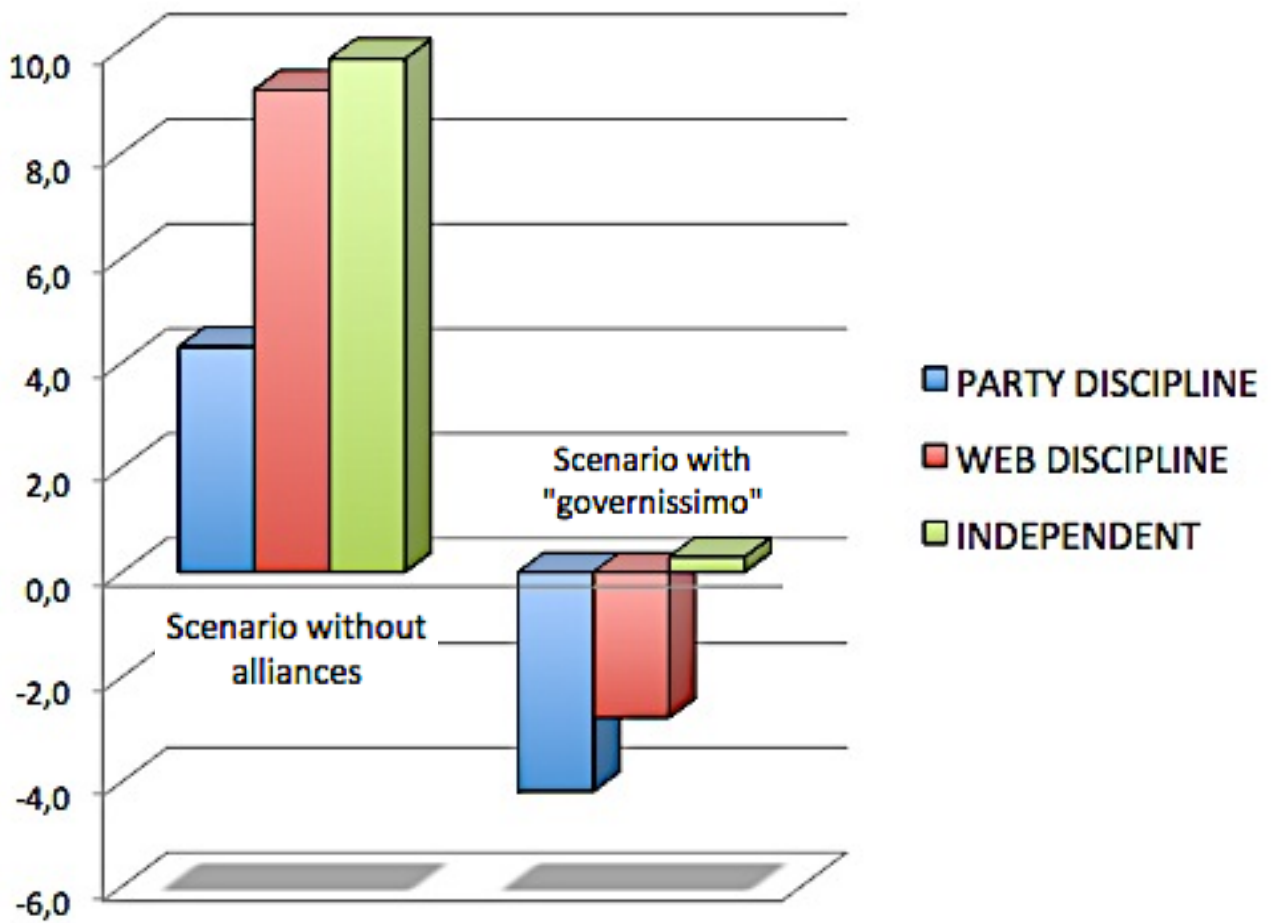

Fig.5

**In conclusion**, within the limits of the working hypotheses and the inevitable simplifications that are at the basis of our model, summing up the results obtained, it is possible to express **a couple of suggestions** for the political forces at play:

1) The **scenario "without alliances" seems to be able to ensure the greatest potential efficiency** for the actual Italian Senate. This possibility appears at first sight in contrast with the common perception of poor governability of a Parliament with variable majorities. However, once the obstacle of the initial trust to some kind of government is passed, this solution is the one that most gives back its **natural function** to the Parliament itself, that is discussing and improving the proposed laws without the "sword of Damocles" imposed by an abuse of the "voting trust" by the government itself (which, usually, is the expression of the coalition possessing the absolute majority).

2) The **role of the 5 Stars Movement** seems to be, in any case, decisive for the purpose of maximizing the efficiency of the Senate. In particular, it seems essential that it would maintain its distance from the traditional parties, especially regarding the voting mode of its senators. In fact, its effectiveness is closely linked to the absence of a fixed center of gravity that collectively influence the vote: the interaction with the electorate via the web or the absence of a binding mandate for senators are **both valid solutions** to have a high efficiency of the system, whatever the location of the senators themselves in the Cipolla's diagram.

## REFERENCES

[1] "Accidental politicians: how randomly selected legislators can improve Parliament efficiency", A. Pluchino, C. Garofalo, A. Rapisarda, S. Spagano, M. Caserta, Physica A 390 (2011) 3944.

[2] "Democrazia a sorte: ovvero la sorte della democrazia", M. Caserta, C. Garofalo, A. Pluchino, A. Rapisarda, S. Spagano, Malcor D' Edizione (2012)

[3] "L'efficienza del caso", A. Pluchino, A. Rapisarda, C. Garofalo, S. Spagano, M. Caserta, Le Scienze, 533 Gennaio (2013) 86.

[4] For more details about our model of the Parliament please visit the web page http://www.pluchino.it/parliament.html

\* \* \*